\documentstyle[times,pramana,epsf,floats]{ias}
\begin{document}
\title{Statistical Analysis of the Road Network of India}

\author{Satyam Mukherjee }
\address{Department of Chemical and Biological Engineering, Northwestern University Evanston, Illinois - 60208}
\keywords{Indian highway network, centrality, community analysis}
\pacs{89.75.Hc}
\abstract{In this paper we study the Indian Highway Network as a complex network where the junction points are considered as nodes, and the links are formed by an existing connection. We explore the topological properties and community structure of the network. We observe that the Indian Highway Network displays small world properties and is assortative in nature. We also identify the most important road-junctions (or cities) in the highway network based on the betweenness centrality of the node. This could help in identifying the potential congestion points in the network. Our study is of practical importance and could provide a novel approach to reduce congestion and improve the performance of the highway network.}
\maketitle
\section{Introduction}
Transportation networks form the backbone of economic development in a country. In the recent years, tools from network analysis have been applied to several transportation networks around the world. These include the airport network of China \cite{cai1}, the airport network of India \cite{bagler}, the world-wide airport network \cite{mossa,barrat}, the urban road networks \cite{holyst} and the railway networks \cite{kim,Sen,cai2,seaton,ghosh}. 
The topology studies of different spatial networks show different degree distributions. Power law degree distribution is seen for Indian airport network \cite{bagler} and world-wide airport network\cite{mossa,barrat}. Also, two regime power laws is observed in China airport network \cite{cai1} and US airport network \cite{cping}. On the other hand exponential degree distribution was observed in railway networks of India \cite{Sen} and China \cite{cai2}.

It is to be noted that several models have been proposed to study various transportation networks. The observed topological properties depend on the way the network is modeled. A directed network model was used to study the Chinese railway network \cite{cai2}. Many transportation networks have  been modeled as bipartite networks and weighted networks \cite{seaton,wangy}. Again, urban networks have been modeled on local optimization process \cite{flammini}. 

Road networks are one of the most prominent means of transport in many countries around the globe and display complex topological properties. For example, urban road networks in Le Mans (France) show double-power law degree distribution \cite{lemans}. The Indian Highway Network (IHN) is one of the busiest road networks in the world, constituting of 2\% of all roads in India but handling 40\% of the total road traffic as of $2010$ \cite{nhai}. The total length of the highways extend up to $70,934$ km. National highways are the backbone of economic development in the country as many cities, towns and industries have come up along the highways. However, according to recent surveys, India faces the challenge of poor rural roads and traffic congestion in major cities. India's road network logistics and bottlenecks in transportation hinder its GDP $1\%$ to $2\%$ (equivalent to US\$16 billion - US\$32 billion).  In this situation it is therefore important to study the topological features of the highway network. Such a study could help in designing efficient modes of extension of the IHN and also a better planning of handling traffic congestion. 	
In this paper we study the Indian highway network as an undirected network. Two nodes(cities/junctions) are connected if there exists a common bus route between them. We measure the basic topological properties of the present IHN and identify potential points of congestion in the network. 

The rest of the paper is laid out as follows. In Section 2 we discuss the data and method of network construction and the topological properties of the network. In Section 3 we have described the community analysis of the Indian highway network. We conclude in Section 4.

\section{Network construction and Topological Properties}
 
\begin {figure}
\epsfxsize=8.0cm
\centerline{\epsfbox{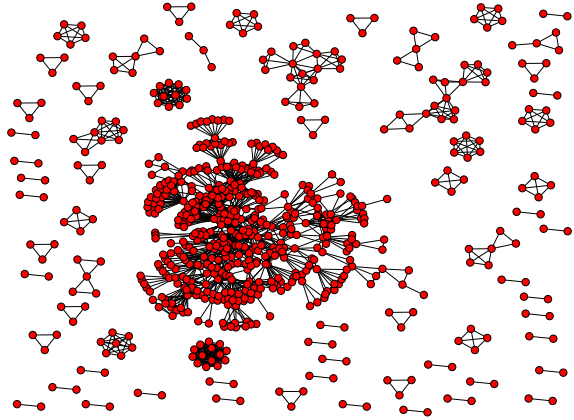}}
\caption{\label{fig:model}(Colour online) Network structure of Indian highways. }
\end{figure}
The highway network is generated as follows. Two cities $b_{i}$ and $b_{j}$ are connected if there exist at least one bus-route which links the two stations. Such a representation has been used earlier in the context of other transportation networks like the railway network \cite{Sen,ghosh} and Indian airport network \cite{bagler}. Figure ~\ref{fig:model} shows the Indian Highway Network (IHN). From the figure it is evident that the IHN has a giant component (which is the largest connected component of the network) and nodes which are not connected to the giant component. This indicates that the IHN is not a fully conneced network. The IHN consist of $694$ nodes and $2209$ edges. 
We analyse the topological properties of the IHN. Clustering coefficient ($C_{i}$) of a node $i$ is defined as the ratio of number of links shared by its neighboring nodes to the maximum number of possible links among them. The average clustering coefficient is defined as,

\begin{equation} 
C = \frac{1}{N}\displaystyle\sum\limits_{i=0}^N C_{i}
\end{equation}
We observe that the average clustering coefficient ($C$) of the network is $0.78$, indicating that the IHN is a highly clustered network. In Table ~\ref{tab:table1} we compare the topological properties of IHN with Erd\"{o}s-R\'{e}nyi model (ER model) \cite{erdos} and Configuration model \cite{configuration}. Keeping the network size same as that of IHN, the network based on configuration model is simulated considering the same degree sequence of IHN.
We observe that the average clustering coefficient ($C$) of the network is $0.78$, indicating that the IHN is a highly clustered network. The clustering coefficient of ER random graph ($0.008$) is an order of magnitude lower than that of the IHN (See Table ~\ref{tab:table1}). We also determine the average shortest path length $D$ between a given vertex and all other vertices of the IHN. 
The average shortest path length is defined as :
\begin{equation} 
D = \displaystyle\sum\limits_{s,t\epsilon V} \frac{d(s,t)}{N(N-1)}
\end{equation}
where $V$ is the set of nodes in the network, $d(s,t)$ is the shortest path from $s$ to $t$, and $N$ is the number of nodes in the network. As shown in Table ~\ref{tab:table1} we observe that $D$ for IHN is of the same order of magnitude as that of ER random graph. These two properties indicate that the IHN is a small world network. 

\begin {figure}
\epsfxsize=8.0cm
\centerline{\epsfbox{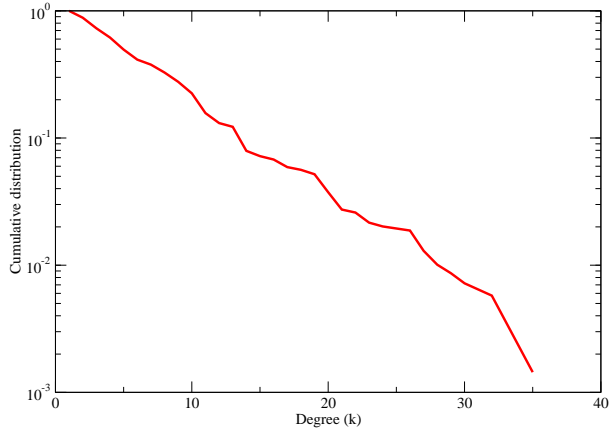}}
\caption{\label{fig:degree}(Colour online) Cumulative degree distribution of the IHN. }
\end{figure}

\begin {figure}
\epsfxsize=8.0cm
\centerline{\epsfbox{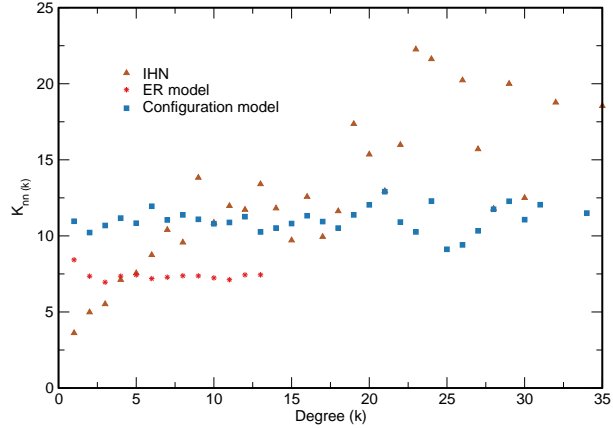}}
\caption{\label{fig:degreeKK}(Colour online) Average unweighted ($K_{nn}(k)$) degree of nearest neighbours of nodes with degree $k$.}
\end{figure}

The degree distribution of a network is defined as the fraction of nodes with degree $k$ in the network. In the case of IHN the degree is defined as the number of cities (junctions) that can be reached from a given city (junction) via a single direct bus. In Figure ~\ref{fig:degree} we show the cumulative degree distribution of the IHN. Another parameter which is studied is the average degree of nearest neighbours, $K_{nn}(k)$ for nodes of degree $k$ (Figure ~\ref{fig:degreeKK}). We observe that $K_{nn}(k)$ increases with degree. This indicates that strong correlations are present among nodes of different degrees. This behaviour is contrary to that observed in the Indian Railway Network \cite{Sen}. The topological assortativity coefficient ($-1 \le r \le 1$), as defined by Newman is another measure of degree correlations in a network \cite{newmanA}. 
Mathematically, its defined as
\begin{equation} 
r = \frac{1}{\sigma_{q}^2}\displaystyle\sum\limits_{jq} jk(e_{jk}-q_{j}q_{k})
\end{equation}
where $q_{k}$ = $\displaystyle\sum_{j}e_{jk}$ and $\sigma_{q}^2$ = $\displaystyle\sum_{k} k^{2}q_{k}$ - ${\displaystyle\sum_{k} kq_{k}}^2$. Here $e_{jk}$ is the probability that a randomly chosen edge has nodes with degree $j$ and $k$ at either end. If the networks are of assortative nature, high degree nodes at one end of a link show some preference towards high degree nodes at the other end. For networks with disassortative mixing, high degree nodes at one end link show preference towards low degree nodes at the other end and vice versa. For the IHN $r=0.46$, indicating a assortative mixing which is consistent with the observation in Figure ~\ref{fig:degreeKK}.

\begin{table}                                                
\caption{\label{tab:table1} Comparison of network properties of Indian Highway Network (IHN) with Erd\"{o}s-R\'{e}nyi model ($p=0.009186$) and Configuration model.}
\begin{tabular}{llll}
$Properties$ & $IHN$&$Erd\text{\"{o}}s-R\text{\'{e}}nyi$&$Configuration$\\
\hline
$v$ & 694 & 694 & 694 \\
$e$ & 2209 & 2148 & 2169 \\
$\bar{k}$ & 6.36 & 6.19 & 6.25 \\
$C$ & 0.78 & 0.008 & 0.015  \\
$D$ & 4.75 & 3.77 & 3.61  \\
\end{tabular}
\end{table}

\begin{table}                                                
\caption{\label{tab:table2} Top $10$ cities of the IHN based on the betweenness centrality (BC).}
\begin{tabular}{cc}
$BC$ & $Junction$\\
\hline
$0.110$ & Delhi \\
$0.107$ & Varanasi\\
$0.081$ & Bangalore \\
$0.051$ & Lucknow \\
$0.050$ & Muzaffarpur \\
$0.040$ & Ranchi\\
$0.03837$ & Raipur \\
$0.03833$ & Mangalore \\
$0.0375$ & Jalandhar \\
$0.0369$ & Chennai \\
\end{tabular}
\end{table}

We also identify the major cities in the IHN. The nodes with high betweenness centrality are potential points of traffic congestion in networks \cite{braj,mukherjee}. Betweenness centrality of a node $v$ is the sum of the fraction of all-pairs shortest paths that pass through $v$:
\begin{equation} 
c_{B}(v) = \displaystyle\sum\limits_{s,t\epsilon V} \frac{\sigma(s,t|v)}{\sigma(s,t)}
\end{equation}
where $V$ is the set of nodes, $\sigma(s,t)$ is the total number of shortest paths and $\sigma(s,t|v)$ is the number os hortest paths passing through $v$ \cite{betweenness}. We list the top ten cities based on the betweenness centrality (See Table ~\ref{tab:table2}). We identify {\it Delhi} as top city according to ranking based on betweenness centrality. This could be {\it posteriori} justified given the fact that {\it Delhi} is the capital city of India and handles maximum traffic. 

\section{Community analysis}

\begin {figure}
\epsfxsize=8.0cm
\centerline{\epsfbox{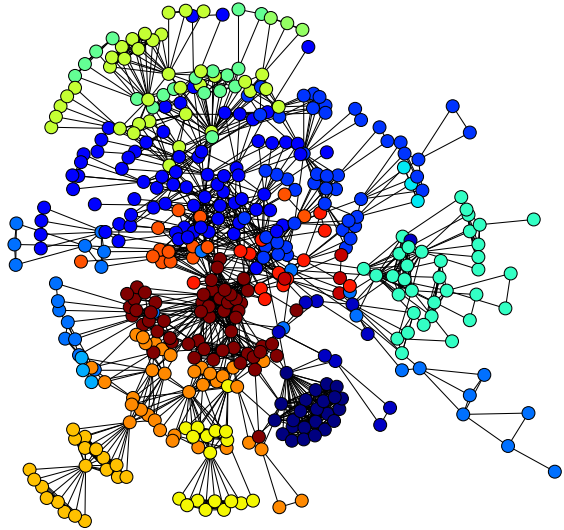}}
\caption{\label{fig:newfig1}(Colour online) Each node indicates a city and each colour corresponds to a community. $17$ communities are identified. Modularity $Q = 0.7718$.  }
\end{figure}
We perform a community analysis on the giant-component of the IHN. To identify the communities in the IHN, we used the definition of modularity introduced by Newman \cite{newmancom}. As described in \cite{newmancom}, modularity is defined as the number of edges falling within groups minus the expected number of edges in an equal sized network of randomly placed edges. Positive values of modularity $Q$, indicates possible presence of community structure. Here we optimize the modularity over possible divisions of the IHN and search for community structure of IHN by looking for positive and large values of $Q$. In Figure ~\ref{fig:newfig1} we display the communities identified by the Newman algorithm in the IHN. We observe that {\it Delhi}, {\it Allahabad}, {\it Lucknow}, {\it Kanpur}, {\it Mathura}, {\it Ludhiana}, {\it Faizabad} and {\it Agra} share the same community. This is evident from the fact that all these cities are located in the Northern parts of India. Interestingly {\it Mumbai} and {\it Dankuni} share the same community with {\it Delhi} even though these two cities are located in the Western and Eastern region of India respectively. Similarly {\it Varanasi} is grouped with the Southern cities like {\it Bangalore}, {\it Mysore}, {\it Hyderabad}, {\it Cochin}, {\it Coimbatore} and {\it Kanyakumari}. This is also justified by the fact that the longest highway in India is {\it NH7} which stretches from {\it Varansi} in north India  to {\it Kanyakumari} in the southern most point of India. We also observe that {\it Chennai} is grouped with {\it Pune}, {\it Belgaum} and {\it Bijapur}. 

\section{Conclusion}
In this paper we have studied the Indian Highway Network (IHN) as an unweighted graph of bus-stations. We observed that the distribution of node-connectivity is neither a power-law nor normal. The IHN shows small world properties and assortative mixing. The cities (junctions) with high betweenness centrality have also been identified. We believe these nodes (cities/junctions) are potential points of congestion. Considering the limited capacity of links to handle road traffic, identification of possible congestion points is of practical importance. With the availability of traffic data, it would also be interesting to study the weighted IHN as it could reveal a clearer picture of network dynamics and the role of cities within a community. The traffic handling capacities of important cities and also that of the links should be enhanced and infrastructure improved in order to achieve efficient traffic.
The IHN is expected to grow in size and achieve maximum connectivity. It would be interesting to study the evolution of the IHN in order to model the topology of the network. Given the availability of traffic data in different cities one could also implement the gravity model on Indian highways as was done for Korean highways \cite{koreanhighway}. It would also be interesting to study the economic growth along the highways and how it is correlated with the centrality of the cities and important junctions. Another possible direction of research is to analyze the road network of metropolitan cities and rural regions, which currently we are unable to study due to unavailability of data. 

\begin{acknowledgments}
We wish to acknowledge the National Highways Authority of India website ($www.nhai.org$) for availability of the data of Indian Highway Network. The author thanks Z. Jabeen for useful comments on manuscript. 
\end{acknowledgments}

\end{document}